\documentclass[useAMS,usenatbib]{mn2e}

\usepackage{epsfig}
\usepackage{subfig}

\newif\ifAMStwofonts
\newcommand{\uc}{ULTRACAM}
\newcommand{\fast}{``fast''}
\newcommand{\slow}{``slow''}
\newcommand{\sloanu}{$\it u'$}
\newcommand{\sloang}{$\it g'$} 
\newcommand{\sloani}{$\it i'$}
\newcommand{\target}{XTE\,J1118+480}

\newcommand{\Msun}{$\rm M_{\odot}$}
\newcommand{\Rsun}{$\rm R_{\odot}$}

\title[\uc\ observations of the black hole X-ray transient 
XTE\,J1118+480 in quiescence.] 
{\uc\ observations of the black hole X-ray transient 
XTE\,J1118+480 in quiescence.}

\author[T.\,Shahbaz et al.]
       {T.\,Shahbaz,$^{1}$\thanks{E-mail: tsh@ll.iac.es}
        V.S.\,Dhillon$^2$,
        T.R.\,Marsh$^3$,
        J.\,Casares$^1$,
        C.\,Zurita$^4$,
        P.A.\,Charles$^5$,
        \newauthor 
        C.\,A.\,Haswell$^6$,
        R.I.\,Hynes$^7$ \\
$^1$Instituto de Astrof\'\i{}sica de Canarias, 38200 La Laguna,
    Tenerife, Spain \\
$^2$Department of Physics and Astronomy, University of Sheffield, 
    Sheffield, S3 7RH, UK  \\
$^3$Department of Physics, University of Warwick, Coventry CV4 7AL, England \\
$^4$Observatorio Astronomico de Lisboa, Tapada da Ajuda 1349-018, 
    Lisboa, Portugal \\
$^5$South African Astronomical Observatory, P.O. Box 9, Observatory, 7935, 
    South Africa\\
$^6$Department of Physics and Astronomy, The Open University, Walton
    Hall, Milton Keynes, MK7 6AA, UK \\
$^7$Department of Physics and Astronomy, Louisiana State University,          
    Baton Rouge, LA 70803-4001, USA   }                

\pagerange{\pageref{firstpage}--\pageref{lastpage}}
\pubyear{2003}

\begin{document}
\maketitle
\begin{abstract}
\noindent

We present high time-resolution multicolour observations of the  quiescent soft
X-ray transient \target\ obtained with \uc. Superimposed on the double-humped
continuum \sloang\ and \sloani-band lightcurves are rapid flare events which
typically last a few minutes.  
The  power density spectrum of the lightcurves  can be described by a
broken power--law model with a break frequency at $\sim$2\,mHz or a  
power--law model plus a broad quasi-periodic oscillation (QPO) at $\sim$2\,mHz. 
In the context of the cellular-automaton we estimate the size of the 
quiescent advection-dominated flow (ADAF) region
to be $\sim 10^{4}$ Schwarzschild radii,
similar to that observed in other quiescent black hole
X-ray transients, suggesting the same underlying physics.
The similarites between the low/hard and quiescent state PDS
suggest a similar origin for the optical and X-ray
variability, most likely from regions at/near the ADAF.

\end{abstract}
%

%%%%%%%%%%%%%%%%%%%%%%%%%%%%%%%%%%%%%%%%%%%%%%%%%%%%%%%%%%%%%%%%%%%%%%%%%%%%
%
% Figure 1
%
\begin{subfigures}
\begin{figure*}
\vspace*{-20mm}
\psfig{angle=0,width=15.0cm,file=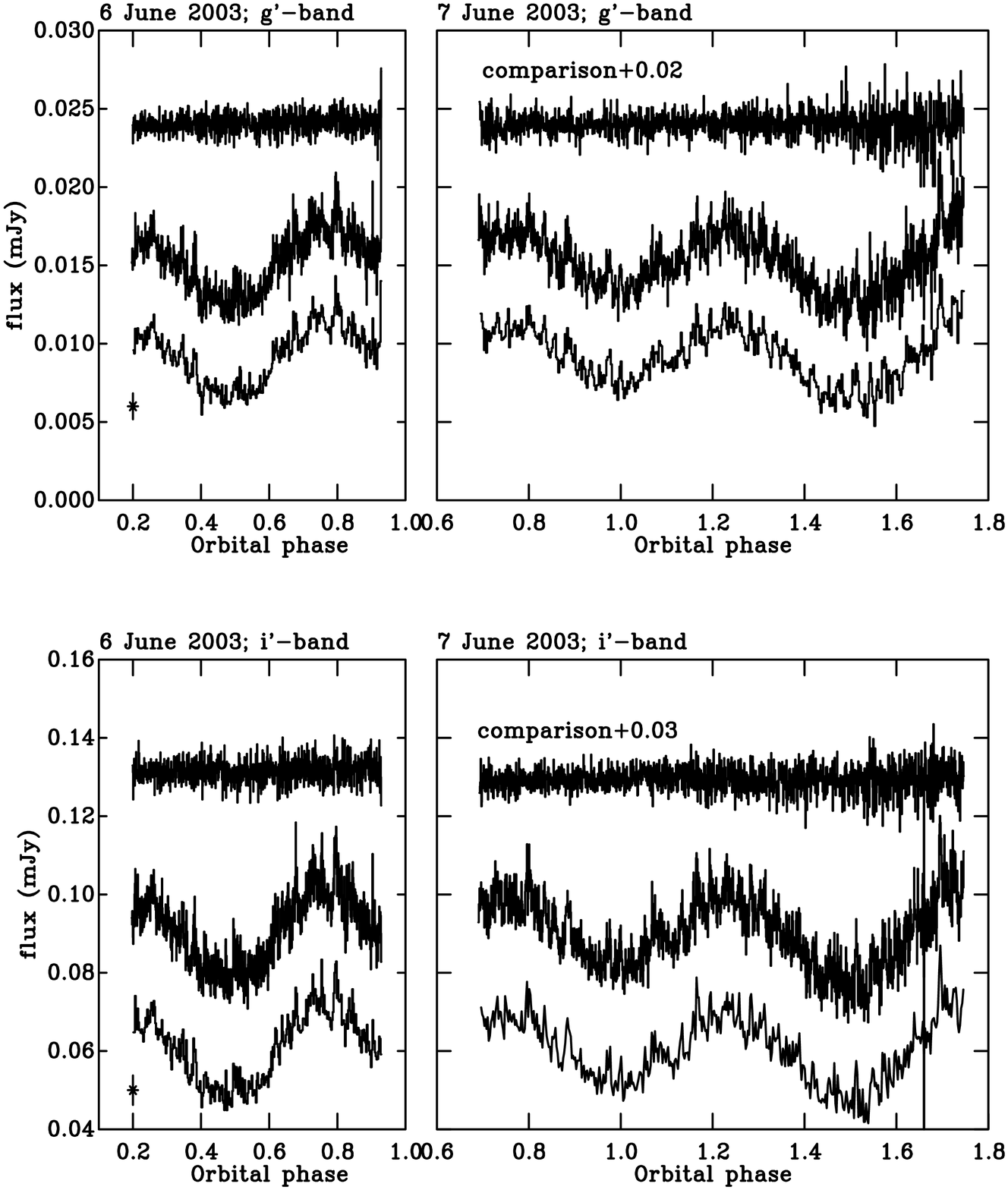}
\caption{The de-reddened \slow\ \sloang\ (top panel)  and \sloani-band 
(bottom
panel)  lightcurves of \target, phase-folded using the ephemeris given in
\citet{McClintock03a}; phase 0.0 is defined as inferior conjunction of the
secondary star. The asterisk marks the typical uncertainty  in the data.
In order to show the short-term flare more clearly, we also show the lightcurve
binned to a time-resolution of 35\,s.  In each panel we also show the
lightcurve of a comparison star of similar  magnitude to \target. }
\label{FIG:SLOWLC}
\end{figure*}
\begin{figure*}
\vspace*{-5mm}
\psfig{angle=0,width=15.0cm,file=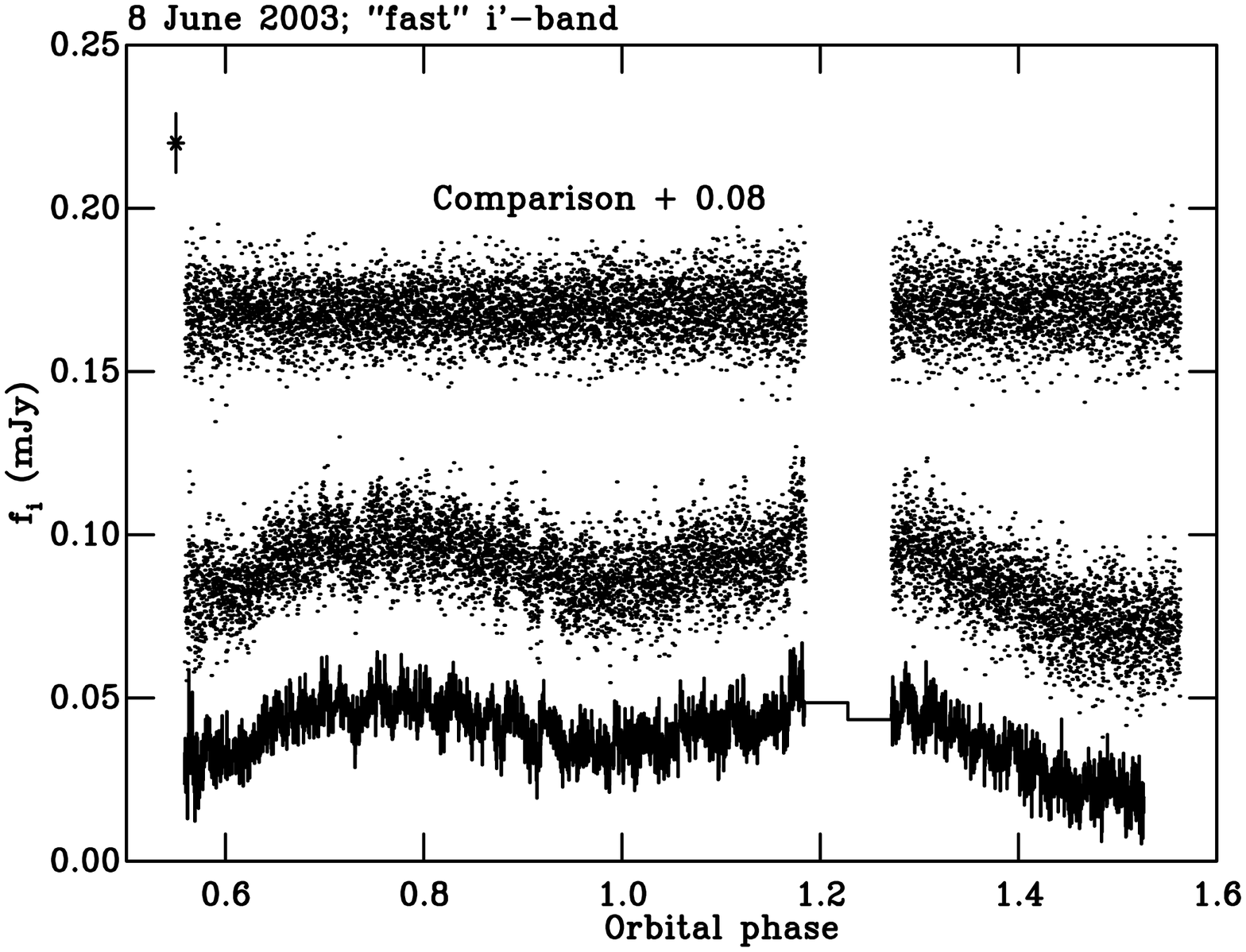}
\vspace*{-10mm}
\caption{The de-reddened phase-folded \fast\ \sloani-band lightcurve 
of \target\ (bottom).
In order to show the short-term flare more clearly, we also show the lightcurve
binned to a time-resolution of 5\,s.
The asterisk marks the typical uncertainty  in the data.
The top lightcurve is of a comparison star of similar  magnitude
to \target. }
\label{FIG:FASTLC}
\end{figure*}
\end{subfigures}

%
%%%%%%%%%%%%%%%%%%%%%%%%%%%%%%%%%%%%%%%%%%%%%%%%%%%%%%%%%%%%%%%%%%%%%%%%%%%%

\section{Introduction}
\label{INTRODUCTION}

X-ray transients (XRTs)  are  a subset of  low-mass X-ray  binaries (LMXBs)
that  display episodic, dramatic X-ray  and optical  outbursts,  usually
lasting for several months. More than 70 percent of XRTs are  thought to
contain black holes \citep{Charles03}.

The XRT \target\ was discovered by the {\it RXTE} All Sky Monitor  on 2000
March 29 \citep{Remillard00}.  Given its very high Galactic latitude   
($\rm b$ = +62 degrees) and correspondingly low interstellar absorption  
($\rm N_{\rm H} \sim 1.2 \times 10^{20}\,cm^{-2}$), it was possible to observe
the multi-wavelength  spectrum including the soft X-ray and extreme
ultraviolet   during outburst (\citealt{Hynes00}; \citealt{McClintock01b}; 
\citealt{Chaty03}). Throughout this time, \target\ remained in  the low/hard
state, characteristic of an accreting black hole binary. 
A power--law spectrum with an index of -1.7  was seen extending out to 120\,keV
(\citealt{Wilson00}; \citealt{Frontera01}) with a similar spectral index to
Cyg\,X--1 in the  low/hard state \citep{Remillard00}.
Faint associated radio emission was detected at 6.2\,mJy  and interpreted in
terms of a steady radio jet \citep{Fender01}. 

%The optical/UV/X-ray spectrum clearly exhibited two components, 
%an X-ray emitting interior region and an exterior region
%which was interpreted as thermal emission from a truncated accretion
%disc (\citealt{Hynes00}; \citealt{Esin01}; \citealt{Chaty03}; 
%\citealt{McClintock03a}). 

A 13th magnitude optical counterpart was promptly identified \citep{Uemura01} 
and its optical spectrum was typical of a black hole X-ray transient in
outburst  \citep{Garcia00}.  A 4.1\,hr weak optical photometric modulation was
discovered  \citep{Cook00}, most likely due to superhumps,  i.e. disc
precession \citep{Uemura02}.  By late November 2000, \target\ was nearly
in quiescence and  optical radial velocity studies led to the determination of the
large binary mass function, {\it f(M)}=6.1$\pm$0.3\,\Msun, thereby establishing
that the compact object is a  black hole (\citealt{Wagner01};
\citealt{McClintock01a};  \citealt{Torres04}).
\target\ has the shortest known orbital period for an XRT of 4.08\,hr
\citep{Zurita02}, the spectral type of the secondary star is approximately K5V 
(\citealt{Wagner01}; \citealt{McClintock01a};  \citealt{McClintock03a}) and the
binary inclination of the system is  high, $i\sim$80\,degrees
(\citealt{Wagner01}; \citealt{Zurita02}).

Black hole X-ray transients are known to exhibit five distinct X-ray  spectral
states, distinguished by the presence or absence of a soft blackbody component
at 1\,keV and the luminosity and spectral slope of emission at harder energies;
the quiescent, low, intermediate, high and very high state \citep{Tanaka96}.
Four of them are successfully explained with the advection dominated accretion
flow (ADAF) model  (Narayan, McClintock \& Yi 1996; Esin, McClintock \& Narayan
1997).  In the context of the ADAF model,   properties similar to the low/hard
state are expected for the  quiescent state, as there is no distinction between
the two except that the mass accretion rate is much higher and the size of the
ADAF region is smaller for the former state.

Unlike the transition between the low/hard and high/soft (thermal-dominant) 
state, where there  is a  reconfiguration of the accretion flow  \citep{Esin97}, 
there is no observational evidence for a transition between the low/hard and
quiescent  states.  In both these states, the ADAF model predicts  that the
inner edge of the disc is truncated at some large radius,  with the interior
region filled by an ADAF. Strong evidence for such a  truncated disc is
provided by observations of \target\ in the low/hard state  during outburst 
(\citealt{Hynes00}; \citealt{McClintock01b}; \citealt{Esin01}; 
\citealt{Chaty03}), where the disc has  an inner radius   of $>$55
Schwarzschild radii ($\rm R_{\rm sch}$) and a hot optically-thin plasma in the
inner regions. With \target\ now in quiescence, the ADAF model predicts that
the inner disc edge will move  outward to larger radii \citep{Esin97}. Indeed,
\citet{McClintock03a} recently fit the X-ray/UV/optical  quiescent spectrum 
of  \target\ and found that it has a hard X-ray spectrum with a  spectral
photon index of $\sim$2, and an optical/UV continuum that resembles a 
13,000\,K blackbody disc spectrum with several strong emission lines 
superimposed.  They presented a two-component accretion flow model,  an
interior region where the flow is advection-dominated  and emits in X-rays, and
an exterior accretion disc truncated at a transition  radius of $\rm R_{\rm
tr}\sim 10^{4}\,R_{\rm sch}$ that is responsible for most of the optical/UV
spectrum.

Here we report on our high-time resolution multi-colour optical  observations
of \target\ in quiescence. We determine the quiescent  PDS, compare it to  the
X-ray low/hard state PDS and determine the size of the ADAF region in
quiescence. 
These observations are part of a continuous campaign with \uc\ to
obtain high-time resolution photometry of X-ray binaries.

\section{Observations and Data Reduction}
\label{OBSERVATION}

Multi-colour photometric observations of \target\ were obtained with \uc\ on
the 4.2-m William Herschel Telescope atop La Palma during the period  2003 June
6 to 8. \uc\ is an ultra-fast, triple-beam CCD camera, where the light is split
into three  broad-band colours (blue, green and red) by two dichroics.  The
detectors are back-illuminated, thinned, E2V frame-transfer  1024$\times$1024
CCDs  with a pixel scale of 0.3\,arcsecs/pixel. Due to the architecture of the
CCDs the dead-time is essentially zero  (for further details see
\citealt{Dhillon01}).

Our observations were taken using the Sloan \sloanu, \sloang\ and \sloani\
filters with effective wavelengths of 3550\,\AA\,, 4750\,\AA\ and 7650\,\AA\
respectively. For the first 2 nights (June 6 to 7), we used an exposure time of
11.65\,s, which was sufficient to give reasonable count rates  in the \sloang\
and \sloani\ bands.   Given the faintness of the object ({\it V}=19.5) and the
short  exposure times, few counts were obtained in the \sloanu\ band.  On the
last  night (June 8) we decreased the exposure time to 1.65\,s,  but only data
in the \sloani\ band were usable.  Hereafter, we will refer to the data taken
with exposure times of 11.65\,s and 1.65\,s as the \slow\ and \fast\ data, 
respectively.

The \uc\ pipeline reduction procedures were used to debias and  flat-field  the
data. The same pipeline was also used to obtain lightcurves for \target\ and
several comparison stars by extracting the counts using aperture photometry. 
The most  reliable results were obtained using a large aperture with a radius
of 1.8\,arcsec.  The count ratio of \target\ with respect to the local
standard  (2.28\,\arcsec\,North 3.75\,\arcsec\,East of \target\ with similar
colour to our target) was then determined.   The magnitude of \target\ was then
obtained using the calibrated magnitude of the local standard. As a check of
the photometry and systematics in the reduction procedure, we also extracted
lightcurves of a comparison star   similar in brightness to the target.  The
mean \sloang\ and \sloani\ band magnitudes of \target\ are 21.00 and 19.04 
respectively and the rms are 14\% and 11\% respectively.
We estimate the  photometric accuracy to be 7.1 and 5.5 percent
for the \slow \sloang\ and \sloani\ band respectively and 6.0 percent  for the
\fast \sloani\ band data.

%%%%%%%%%%%%%%%%%%%%%%%%%%%%%%%%%%%%%%%%%%%%%%%%%%%%%%%%%%%%%%%%%%%%%%%%%%%%
%
% Figure 2
%
\begin{figure}
\psfig{angle=0,width=8.5cm,file=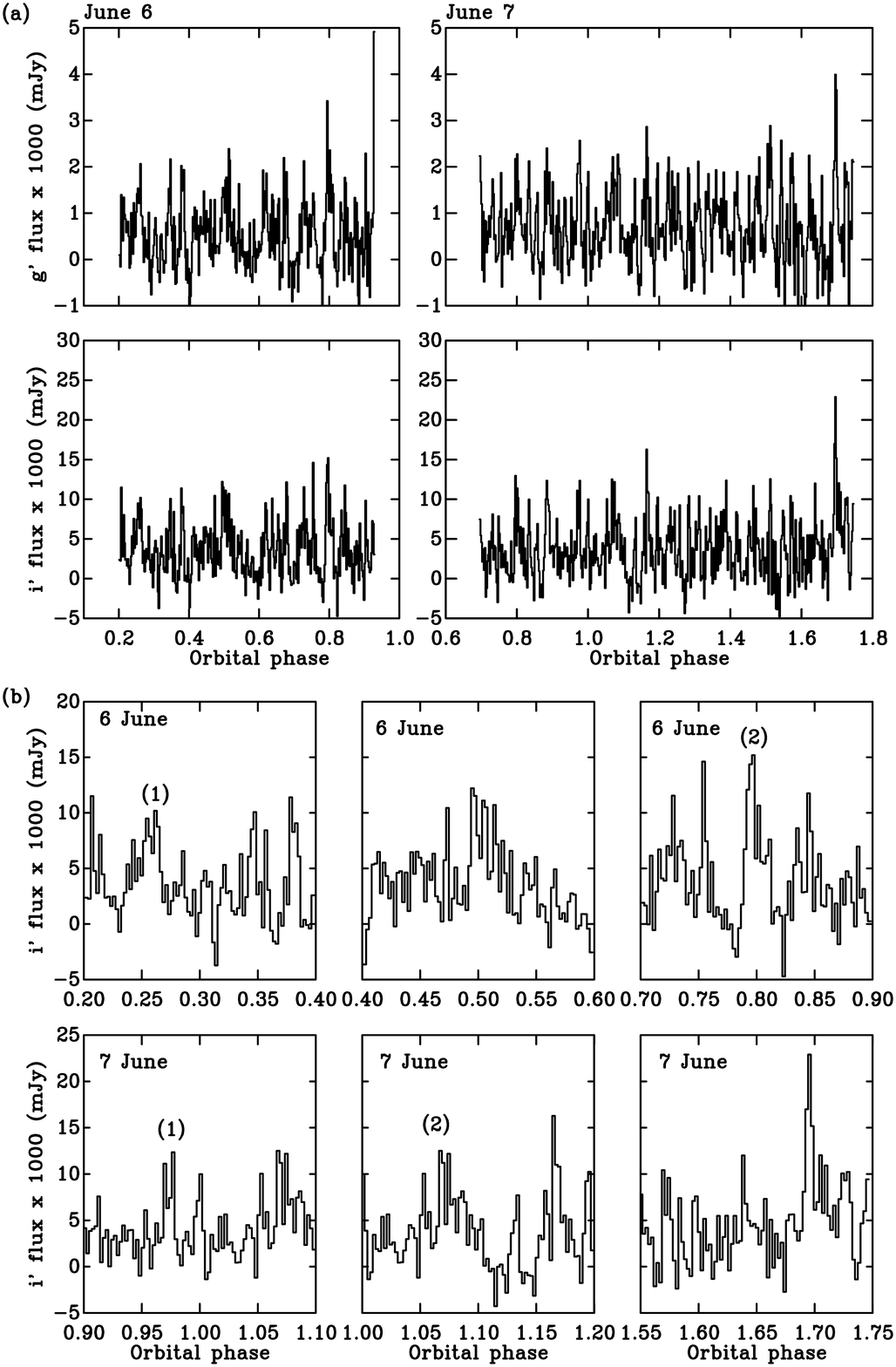}
\caption{(a) The flare de-reddened flux density \sloang\ (top panel) and
\sloani-band (bottom panel) lightcurves of \target\ obtained by subtracting a
fit to the lower-envelope of the lightcurves in Figure\,1. 
For clarity, the flare
lightcurves  have been binned by a factor of 3 (i.e. a time-resolution of 
35\,s).  The
uncertainties in the \sloang\ and \sloani\ binned lightcurves are  3.6$\times
10^{-4}$\,mJy and 1.3$\times 10^{-3}$\,mJy, respectively. (b) Detailed plots of
some individual flares in the  \sloani-band lightcurve. }
\label{FIG:FLARES}
\end{figure}
%
%%%%%%%%%%%%%%%%%%%%%%%%%%%%%%%%%%%%%%%%%%%%%%%%%%%%%%%%%%%%%%%%%%%%%%%%%%%%

%%%%%%%%%%%%%%%%%%%%%%%%%%%%%%%%%%%%%%%%%%%%%%%%%%%%%%%%%%%%%%%%%%%%%%%%%%%%
%
% Figure 3
%
\begin{figure}
\vspace{10mm}
\hspace{-5mm}
\psfig{angle=0,width=9.0cm,file=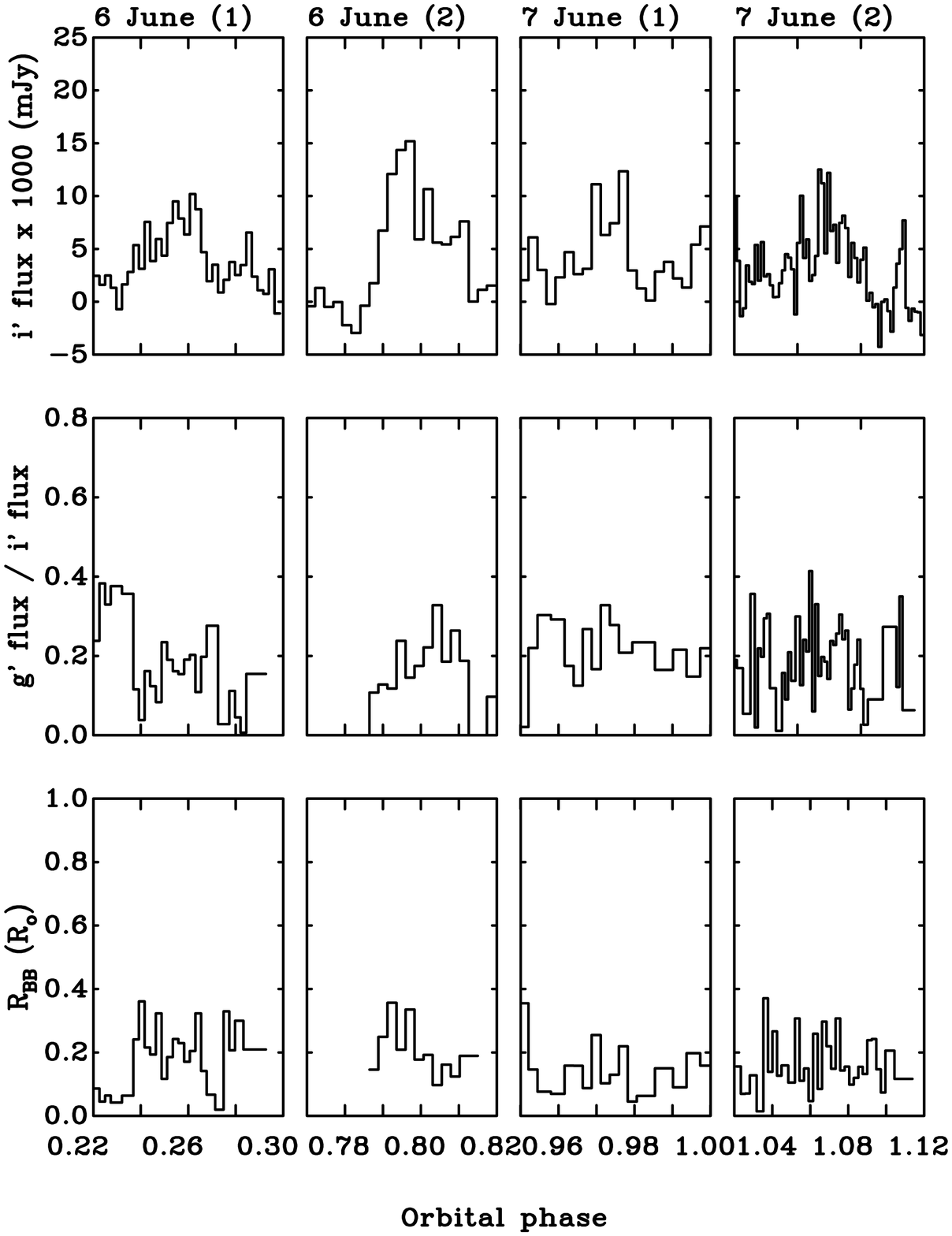}
\caption{ 
In the top panel we show some \sloani-band flare profiles shown in 
Figure\,\ref{FIG:FLARES}(b), re-binned by a factor of 3 for clarity. 
The middle panel shows the flux density ratio $\rm f_{g'}/f_{i'}$ and the
bottom panel shows the projected blackbody radius of the region producing the
flares.}
\label{FIG:FRATIO}
\end{figure}
%
%%%%%%%%%%%%%%%%%%%%%%%%%%%%%%%%%%%%%%%%%%%%%%%%%%%%%%%%%%%%%%%%%%%%%%%%%%%%

\begin{table}
\caption{Properties of the flares. $v_{obs}$ is the spectroscopic veiling,
$v^{'}_{d}$ is the contribution to the  {\it non-variable} disc light,
$\bar{z}_{f}$ and $\sigma_{z}$ are the mean flare flux and its standard
deviation respectively.  $\sigma^{*}_z = \sigma_z / v^{'}_{d}$  and  $\eta$ is the fraction
of the average veiling due to the flares \citep{Zurita03}.
}
\label{Table:Flares}
\begin{center}
\begin{tabular}{lcccccc}
\hline
band    & $v_{obs}$      & $v^{'}_{d}$ & $\bar{z}_{f}$       & $\sigma_z$   
        & $\sigma^{*}_z$ & $\eta$        \\
\noalign{\smallskip}
\sloang & 40\%$^1$       &   39.9\%     & 0.0007 &  0.0011   
        &  0.0027        &    0.2\%      \\ 
\sloani &  -             &     -         & 0.0036 &  0.0050
        &    -           &     -         \\ 
\noalign{\smallskip}
\hline
\end{tabular}
\end{center}
$^1$ \citet{Torres04}
\end{table}

\section{The short-term variability}
\label{VAR}

It is clear that the optical lightcurve of \target\  
(Figure\,\ref{FIG:SLOWLC}, \ref{FIG:FASTLC}) is dominated by  the 
secondary star's ellipsoidal,  which is not linked to the short-term
variability/flares  (Zurita, Casares \& Shahbaz 2003, \citealt{Hynes03a},
\citealt{Shahbaz03b}). Therefore if we want to determine the flux of the
flares, these ``steady'' contributions  must first be removed from the
lightcurves. 
In order to isolate the short-term variability in each band. we first
de-reddened the observed magnitudes using a colour excess of E(B-V)=0.017
\citep{Chaty03}  and the ratio  $A_{\rm V}/E(B-V)$=3.1  \citep{Cardelli89}, 
giving \sloang\ and \sloani\ extinction values of 0.06 and 0.04 mags respectively
and then converted the Sloan AB magnitudes to flux density \citep{Fukugita96}.
We  then fitted a double sinusoid to the  lower-envelope of the   lightcurve 
with periods equal to the orbital  period and its first harmonic, where the
phasing was allowed to float free. We rejected points more than 3-$\sigma$
above the fit, then refitted,  repeating the procedure until no new points were
rejected  (\citealt{Zurita03}; \citealt{Hynes03a}). 
The resulting lightcurves did not show any long-term structure,
suggesting  that any contamination from a superhump modulation is weak; at the
$<$0.50\% level.
As one can see from Figure\,\ref{FIG:FLARES}, there are numerous rapid flare 
events, which typically last 5\,min or less. 
The parameters of the flares as defined by \citet{Zurita03} are given in 
Table\,\ref{Table:Flares}. 
Using these flare  lightcurves we
determined the flux density ratio $\rm f_{g'}/f_{i'}$ and the  equivalent
blackbody radius $\rm R_{\rm BB}$. To determine the  colour temperature 
corresponding to a given flux density ratio, we integrated blackbody functions 
with the CCD and Sloan filter response functions  and then determined the  $\rm
f_{g'}/f_{i'}$ flux density  ratio. Given this blackbody temperature we  can
then determine the  corresponding radius of the region that produces the
observed de-reddened flux at a distance of 1.7\,kpc \citep{Chaty03}. 
Typically, the flares have $\rm f_{g'}/f_{i'} \sim 0.20$, similar to the flares
in V404\,Cyg \citep{Shahbaz03b} and $\rm R_{\rm BB}\sim 0.10\,R_{\odot}$. (see
Figure\,\ref{FIG:FRATIO}).  Following the method described in
\citet{Shahbaz03b}, we estimate  the ratio  $\rm f_{g'}/f_{i'} \sim 0.20\pm
0.10 $  corresponds to a LTE slab of hydrogen with a temperature  of 
$\sim$3500$\pm$500\,K.

%%%%%%%%%%%%%%%%%%%%%%%%%%%%%%%%%%%%%%%%%%%%%%%%%%%%%%%%%%%%%%%%%%%%%%%%%%%%
%
% Figure 4
%
\begin{figure}
\hspace*{5mm}
\psfig{angle=90,width=12.5cm,file=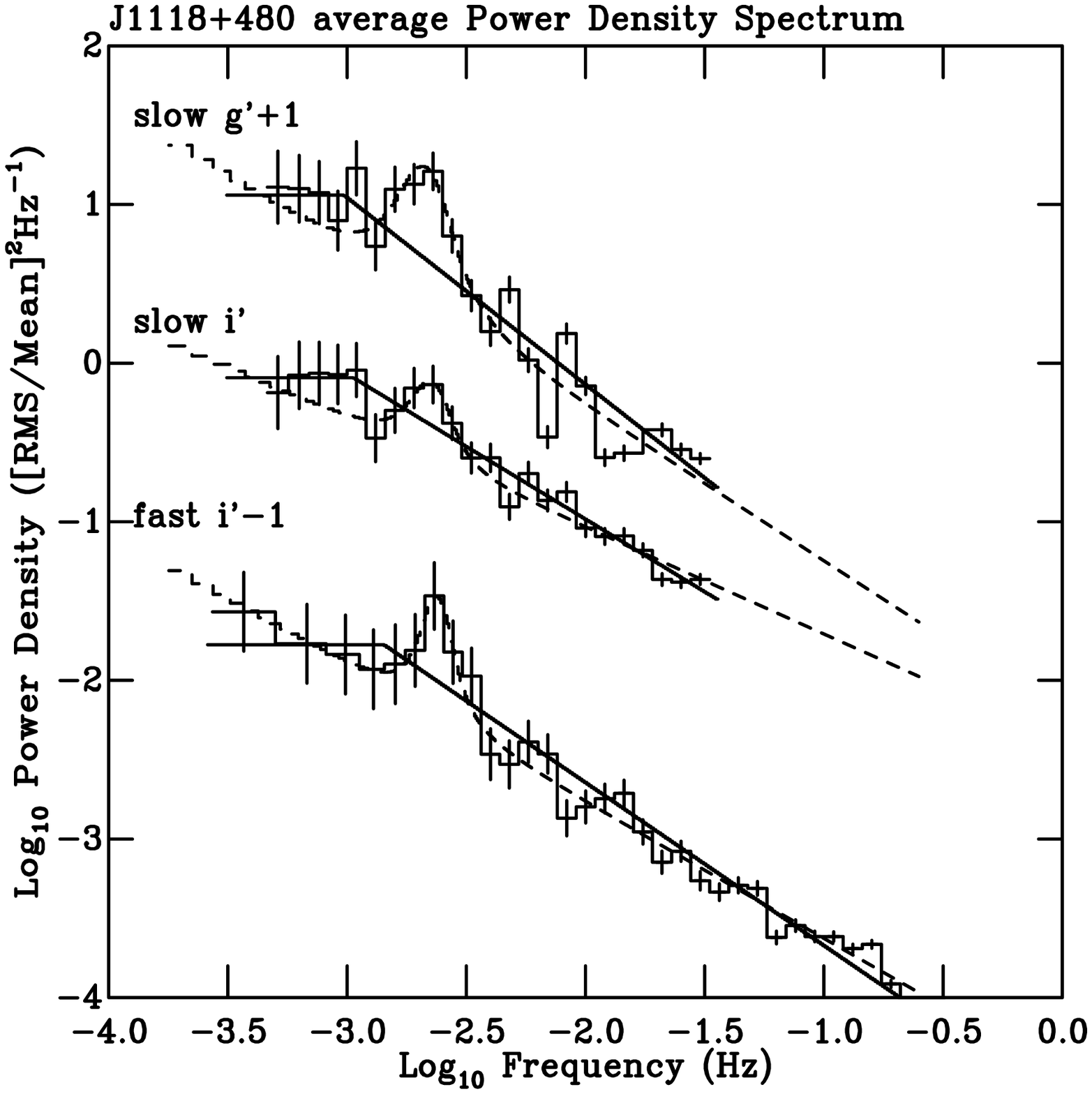}
\caption{The PDS of \target. 
From top to bottom: \slow\ \sloang\ and \sloani\ PDS, and the 
\fast\ \sloani\ PDS. There is a vertical offset between the spectra.
The solid line is a fit with a broken power--law and the
dashed line is a fit with a power--law+QPO model.
A QPO at $\sim$2\,mHz is noticeable in all the spectra.}
\label{FIG:PDS_OP}
\end{figure}
%
%%%%%%%%%%%%%%%%%%%%%%%%%%%%%%%%%%%%%%%%%%%%%%%%%%%%%%%%%%%%%%%%%%%%%%%%%%%%

\section{The power density spectrum}
\label{PDS}

To compute the power density spectrum (PDS) of the \slow\ and \fast\ data,  we
detrended the data using the double sinusoid fit described in the previous
section and then added the mean  flux level of the data.   Although the \uc\
sampling is perfectly uniform, we use  the Lomb-Scargle method  to compute the 
periodograms \citep{Press92} with the same normalization method as is commonly
used in  X-ray astronomy, where the power is normalized to the fractional root
mean amplitude squared per hertz \citep{Klis94}.  We used the constraints
imposed by the Nyquist frequency  and the typical duration of each observation
and binned and fitted the PDS in logarithmic space \citep{Papadakis93}, where
the errors in each bin are  determined from the standard deviation of the
points within each bin.  The  white noise level was subtracted by fitting the
highest frequencies with a  white-noise (constant) plus red-noise (power--law)
model. 

%%%%%%%%%%%%%%%%%%%%%%%%%%%%%%%%%%%%%%%%%%%%%%%%%%%%%%%%%%%%%%%%%%%%%%%%%%%%
%
\begin{table*}
\caption{Fitted properties of the optical PDS for \target.}
\label{Table:PDS}
\begin{center}
\begin{tabular}{lcccccc}
\hline
\noalign{\smallskip}
PDS   &  Slope &    Break-freq.     &  QPO   & FWHM  & rms &  Q  \\
      &        &  (mHz) & (mHz)   & (mHz) & (\%) &   \\
\hline
\slow\ \sloani         &                &              &       &     &    &    \\
\hspace*{4mm} PLB      &  1.8$\pm$0.2   & 1.9$\pm$0.5  &  -    & -   & -  &  - \\
\hspace*{4mm} PL+QPO   &  1.4$\pm$0.2   & -     &  2.2$\pm$0.3  & 0.7$\pm$0.8 
                       &  3.1 & 3.1 \\ 
\slow\ \sloang         &                &              &       &     &    &   \\
\hspace*{4mm} PLB      &  2.9$\pm$0.3   & 1.0$\pm$0.7 &  -    & -   & -   & - \\
\hspace*{4mm} PL+QPO   &  1.9$\pm$0.2   & -   &  2.1$\pm$0.2  & 0.8$\pm$0.3 
                       &  5.4 & 2.6 \\
\fast\ \sloani         &                &              &       &     &    &    \\
\hspace*{4mm} PLB      &  2.3$\pm$0.2   & 1.9$\pm$0.3  &  -    & -   & -  &  - \\
\hspace*{4mm} PL+QPO   &  1.7$\pm$0.2   & -    &  2.4$\pm$0.4  & 0.5$\pm$1.3 
                       &  1.9 & 4.8 \\
\noalign{\smallskip}
\hline
\end{tabular}
\end{center}
\begin{tabular}{l}
$^*$PL refers to power--law fit of the form $P\propto\nu^{\alpha}$; \\
PLB refers to a power--law break and broad QPO refers to a 
quasi-periodic oscillation; \\ 
''rms'' refers to the fractional root-mean squared amplitude;
''Q'' refers to the quality factor.
\end{tabular}
\end{table*}
%
%%%%%%%%%%%%%%%%%%%%%%%%%%%%%%%%%%%%%%%%%%%%%%%%%%%%%%%%%%%%%%%%%%%%%%%%%%%%

In Figure\,\ref{FIG:PDS_OP}, we show the PDS which can be described by a broken
power--law model or a power--law model + quasi-periodic  oscillation (QPO). 
One can see suggestions of a broad QPO in all the PDS, at $\sim$2\,mHz and/or a
break at $\sim$2\,mHz.  Note that the \slow \sloang\ and \sloani\ PDS are not
independent,  being simultaneous, but the \fast \sloani\ is independent, and
shows  the same (even stronger) feature.   Table\,\ref{Table:PDS} gives the
fitted properties of the PDS. The Q-factor, defined as the centroid frequency
divided  by the FWHM of the peak in the PDS. Given that the QPO has Q$>$3, in
the following sections we will refer to the  possible QPO as a "broad QPO"	

\begin{figure}
\hspace*{10mm}
\psfig{angle=90,height=8.5cm,file=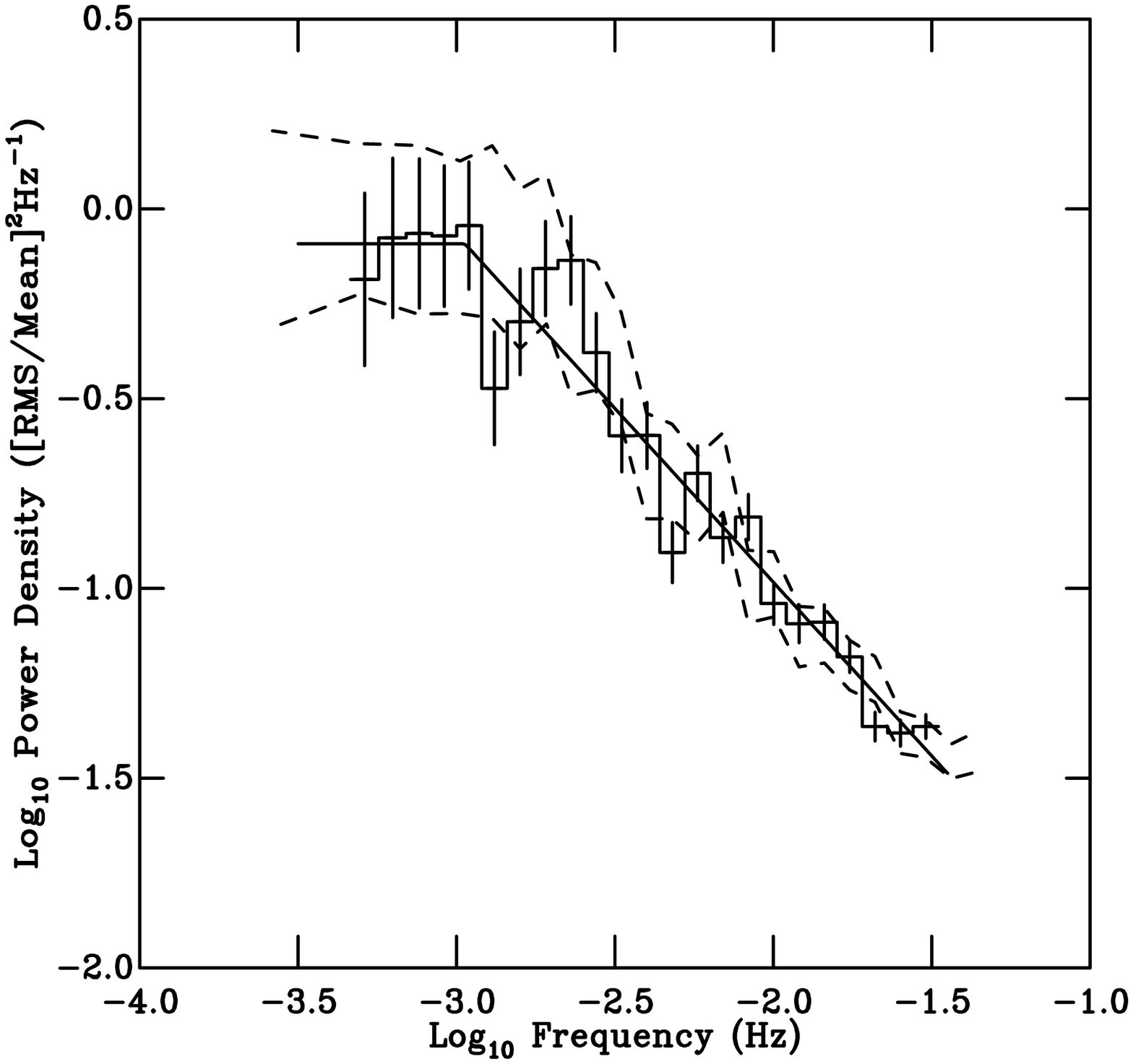}
\hspace*{10mm}
\psfig{angle=90,height=8.5cm,file=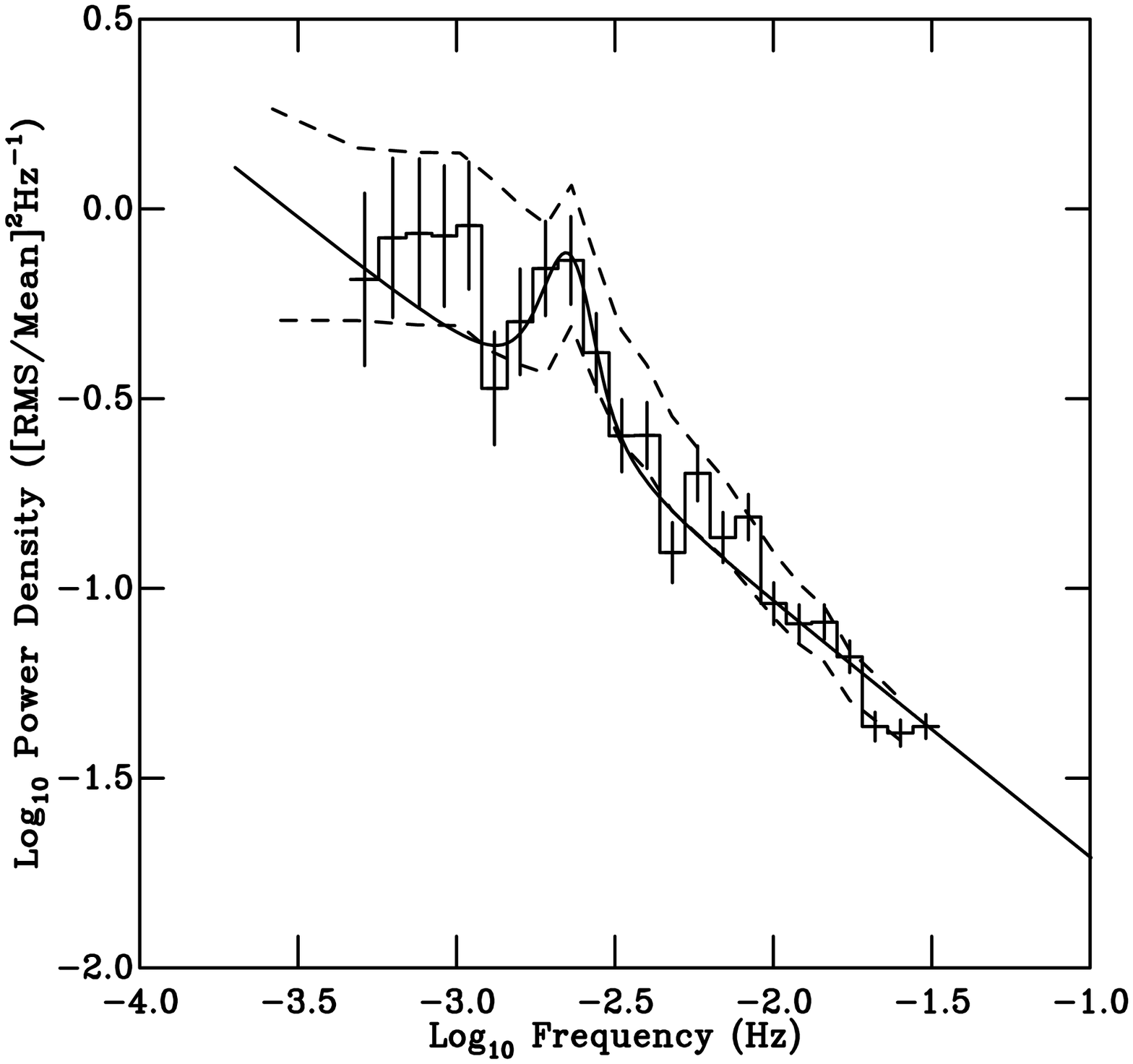}
\caption{Top: The results of the Monte Carlo simulation for a PDS
consisting of a  broken power--law model. The dotted lines show the
1-$\sigma$  confidence region for individual points obtained by computing 1000
simulations (see text). 
Bottom: same as above but for a power--law + broad QPO model.
}
\label{FIG:PDS_MCQPO}
\end{figure}

We tested the significance of the models to describe the PDS using a  Monte
Carlo simulation similar to \citet{Hynes03a}. We generated lightcurves with
exactly the same sampling and integration times as the real data. We started
with a model for the ellipsoidal modulation,  calculated using the X-ray binary
model described in \citet{Shahbaz03a}  with the parameters given in
section\,\ref{VAR}.    To this we added a model noise lightcurve  generated
using a power--law index of --1.0, a break-frequency at 1\,mHz, calculated
using the method of \citet{Timmer95}, or  a broad QPO at 2\,mHz with similar strength
as observed in the \slow\ PDS.   We then added Gaussian noise using the errors
derived from the  photometry. We calculated 1000 simulated lightcurves and
analyzed them in exactly the same way as for the real data. We created
individual PDS with the same logarithmic frequency binning used for the data. 
With the 1-$\sigma$ confidence levels, we found that the   broken power--law
and the power--law + QPO model produced a PDS that matched  the  observations
equally (see Figure\,\ref{FIG:PDS_MCQPO}). We therefore conclude that given the
uncertainties in the data, the PDS can be described by either the broken
power--law or a power--law + broad QPO model.

%%%%%%%%%%%%%%%%%%%%%%%%%%%%%%%%%%%%%%%%%%%%%%%%%%%%%%%%%%%%%%%%%%%%%%%%%%%%
%
% Figure 5
%

%
%%%%%%%%%%%%%%%%%%%%%%%%%%%%%%%%%%%%%%%%%%%%%%%%%%%%%%%%%%%%%%%%%%%%%%%%%%%%

\section{The break frequency}
\label{BMODELS}

Most models for the observed optical and X-ray variability from an accretion
disc around a black hole [see \citet{Wallinder92} for a review] can produce
periodic or quasi-periodic time variations.  To reproduce the observed X-ray
fluctuations, the physical quantities of the disc must change abruptly and in
order to explain the  observed 1/f-like fluctuations, a smooth distribution  of
flares on a variety of time-scales is required. Here we outline the two most
attractive mechanisms to produce the 1/f-like fluctuations in X-ray binaries
and cataclysmic variables.

\subsection{The self-organized criticality model}
\label{SOC}

\citet{Mineshige94a} proposed a cellular-automaton model using the concept of
self-organized criticality of \citet{Bak88} to explain the observed 1/f-like
X-ray fluctuations in X-ray binaries.  In their model, the accretion disc
comprises of two parts: an outer disc where the disc material smoothly drifts
inwards, and an inner disc which suffers an instability. Gas particles are
randomly injected into the inner regions of the accretion disc around a black
hole. The system then evolves to and stays in a self-organized critical state
in spite of random mass injections. Mass accretion occurs either by an
avalanche, which is triggered when the mass density of the disc exceeds some
critical value, or by gradual gas diffusion \citep{Mineshige94b}.  

\citet{Manmoto96} and \citet{Takeuchi97} took this work further and 
investigated the response of an advection-dominated disc to an assumed thermal
perturbation, producing models in which such  X-ray shots could be 
continuously created.   Since the standard-disc cannot explain the long
time-scale of the  observed X-ray fluctuations,  an advection-dominated disc is
favoured, since it is characterized by low-emissivity and a large infall
velocity.  The main energy release occurs by sporadic reconnection events  
\citep{Haswell92},  leading to magnetic eruptions, similar to solar flares.
Magnetic energy is stored and released in an avalanche. They found that the
optically thin solutions, dominated by advection (\citealt{Ichimaru77};
\citealt{Narayan94}),  could produce persistent, but fluctuating hard X-ray
emission.  Assuming that a disturbance is produced by some critical behavior in
the disc, they successfully reproduce the 1/f-like fluctuations at high
frequencies and also the break in the PDS at low frequencies, as seen in the
PDS of X-ray binaries.  The range in frequency over which the 1/f-like
fluctuations are observed, depends on the size of the ADAF region of the disc.
Beyond the ADAF region (in the outer disc regions),  matter accretes smoothly
via the usual viscous diffusion process, thus producing white-noise.

The break-frequency is determined by the maximum peak intensity of the X-ray 
shots which is on the order of the size of the advection-dominated region,  and
corresponds to  the inverse of the free-fall time-scale of the largest
avalanches (Takeuchi, Mineshige \& Negoro 1995)  at the radius of the ADAF
region ($\rm R_{\rm crit}$). As shown in equation 13 of \citet{Takeuchi95} 

\begin{equation}
\frac{R_{\rm crit}}{R_{\rm sch}} \sim 10^{3.2} \left( \frac{f_{\rm break}}{0.1} \right)^{-2/3}
\left( \frac{M_{\rm X}}{10M_{\odot}} \right)^{-2/3}
\end{equation}

\noindent 
where $\rm R_{\rm sch} = 2GM_{\rm X}/c^{2}$ is the Schwarzschild radius for a
black hole with mass $\rm M_{\rm X}$. However, it should be noted that the
break frequency depends not only on the size of the ADAF region but also on the
propagation speed of the perturbation (Mineshige priv. comm.). Since the
perturbation velocity should be less than the  free-fall velocity, the 
free-fall velocity gives an upper limit to the size of the ADAF. Using 
$\rm M_{\rm X}$=7.2\,\Msun and  the $\rm f_{\rm break}\sim$2\,mHz observed in
quiescence (section\,\ref{PDS}), we find 
$\rm R_{\rm crit} < 3 \times 10^{4}\,R_{\rm sch}$ (=0.9\,\Rsun).

\subsection{The fluctuating viscosity model}
\label{VISCOSITY}

\citet{Lyubarskii97} considered an ADAF disc and showed that the 1/f-like
fluctuations in the accretion rate (and thus luminosity) near the inner radius
of the disc can in principle be caused by the viscosity fluctuating 
independently (most likely due to a magnetic dynamo) at different radii on the
local viscous time-scale. \citet{Lyubarskii97} was able to reproduce a
power--law PDS with a frequency break given by the Keplerian accretion
time-scale  $\rm \tau = (\alpha \Omega_{\rm K}^{-1}$),  where  $\Omega_{\rm K}$
is the Keplerian angular velocity and  $\alpha$ is the dimensionless viscosity
parameter \citep{Shakura73}.  For an observed break-frequency of 2\,mHz, this
corresponds to an ADAF with an outer disc radius of 
$\rm \sim 10^{4}\,R_{\rm sch}$.

Recently, \citet{King04} have shown how the local magnetic dynamo in the disc
can  also affect the disc evolution. Angular momentum losses due to a disc
wind/jet can significantly drive the accretion rate in systems which are in a
jet-dominated state \citep{Fender03}.  \citet{King04} consider an accretion
disc in which a magnetic dynamo generates the viscosity, but also occasionally
produces a well-ordered poloidal field which  affects the accretion rate by
driving a wind/jet \citep{Livio03}.  The luminosity arises from  viscous
dissipation in the disc after allowing for the energy removed by a wind/jet, 
and produces a  1/f-like PDS and a break-frequency  given by the magnetic
alignment time-scale at the inner disc edge. However, it should be noted that
in their models the disc extends to  the last stable orbit around a black hole,
as is the case for the X-ray high/soft thermal dominant  state.  Although this
is an attractive model, because in principle the same model could be used to
explain the flickering in cataclysmic variables, it is clear that it needs to
be  adapted to include an ADAF in the low/hard or quiescent states.

%%%%%%%%%%%%%%%%%%%%%%%%%%%%%%%%%%%%%%%%%%%%%%%%%%%%%%%%%%%%%%%%%%%%%%%%%%%%
%
% Figure 6
%
\begin{figure}
\hspace{-5mm}
\psfig{angle=0,width=12.5cm,file=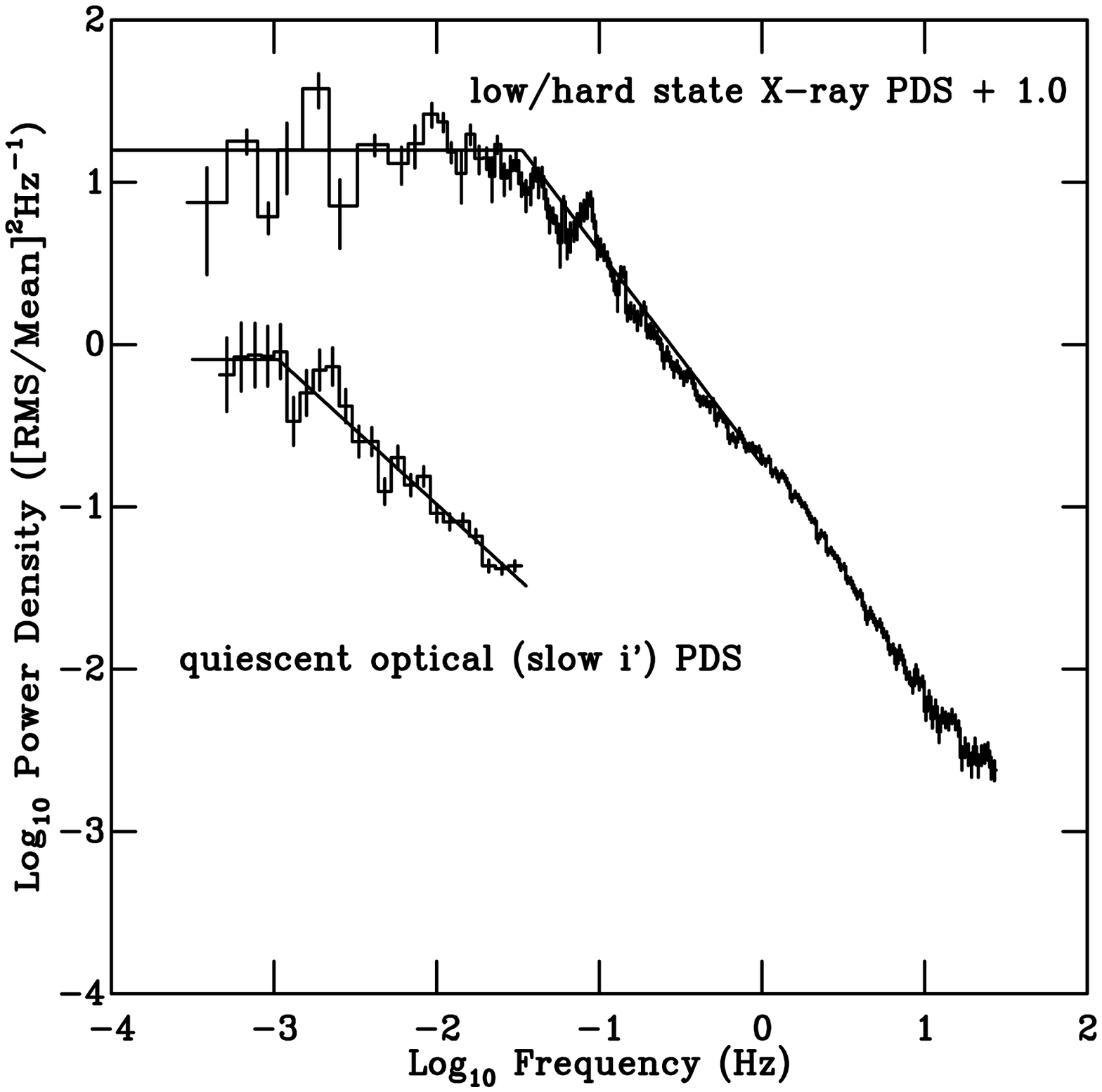}
\caption{The quiescent optical (\slow\ \sloani) and low/hard state X-ray PDS of
\target\ (taken from \citealt{Hynes03b}). Note the shift in the break-frequency
and broad QPO.
The solid lines are broken power--law fits.}
\label{FIG:PDS_OPX}
\end{figure}
%
%%%%%%%%%%%%%%%%%%%%%%%%%%%%%%%%%%%%%%%%%%%%%%%%%%%%%%%%%%%%%%%%%%%%%%%%%%%%

\section{The low-frequency $\sim$2\,mHz broad QPO}
\label{LFQPO}

During the outburst of \target\ a $\sim$0.08\,Hz QPO was reported in the
X-rays  \citep{Revnivtsev00} and subsequently confirmed by {\it ASCA} data 
\citep{Yamaoka00} as well as {\it RXTE} observations \citep{Wood00}.
Simultaneous {\it RXTE/HST} observations showed a similar QPO \citep{Haswell00}
and that the  variability in the  X-rays and optical/UV bands were correlated
\citep{Hynes03b}. In Figure\,\ref{FIG:PDS_OP} there is a suggestion of 
a broad QPO in the quiescent simultaneous \slow\ \sloang\ and \sloani\ PDS 
{\it and} the independent \fast\ \sloani\  PDS at $\sim$2\,mHz.

In the context of a two-component accretion flow model (\citealt{Narayan96};
Narayan, Barret \& McClintock 1997)  an ADAF has turbulent gas at all radii, with a variety of
time-scales, ranging from a slow time-scale at the transition radius down to
nearly the free-fall time close to the black hole.  Interactions between the
hot inner ADAF and the cool, outer disc,  at or near the transition radius, 
can be a source of optical QPO variability, due to synchrotron emission by the
hot electrons in the ADAF, and would have a characteristic time-scale given  
by a multiple of the Keplerian  rotation period at the transition radius. If we
assume that the 2\,mHz broad QPO observed in the PDS is the  dynamical time-scale at
the transition radius, then the transition radius  lies at 
$\sim$8000\,$\rm R_{\rm sch}$.

\section{Discussion}
\label{DISCUSSION}

Band-limited noise and QPO's are a common feature in the low/hard state  XRTs
and many LMXBs \citep{McClintock03b}. Indeed, observations of \target\ in the
low/hard state reveal the presence of  band-limited noise at X-ray and
optical/UV wavelengths \citep{Hynes03b}. For comparison, in
Figure\,\ref{FIG:PDS_OPX} we show the low/hard state X-ray PDS of \target\
along with our optical  quiescent PDS (section\,\ref{PDS}). The low/hard state
X-ray PDS shows a low-frequency break at 23\,mHz and a QPO at $\sim$80\,mHz,
whereas  the optical quiescent PDS shows either a break at a much lower
frequency  of $\sim$ 2\,mHz or a broad QPO at $\sim$2\,mHz.  
It  is interesting to compare the optical and X-ray PDS assuming that the
optical PDS can be described by either a break-frequency or a broad QPO model. 
The position of the quiescent optical broad QPO is a factor of  $\sim$40 lower
than the low/hard state QPO, and the quiescent optical break-frequency is 
$\sim$12 lower than the low/hard state break-frequency. The presence of the
possible break-frequency or a broad QPO (possibly with a multiple frequency
ratio)  provides evidence that we are seeing the same phenomenon in outburst.
%and  that the low/hard and quiescent states are analogous. 
The similarity between the low/hard and quiescent state  PDS suggest that the
optical variability could have a similar origin to the X-ray PDS and might be
associated with the size of the direct emission from the self-absorbed synchrotron emission
arising from an advective-dominated flow (see \citealt{Narayan94} and
references therein) or from optically thin synchrotron emission directly from a
jet (\citealt{Markoff01b}; \citealt{Fender03}).
The quiescent state would have a larger ADAF region than in the low/hard
state.

The slope of the \sloang\ PDS is steeper than the slope of the \sloani\ PDS,
in  contrast to what is observed in V404\,Cyg \citep{Shahbaz03b}. This
difference could  be due to either more low-frequency disc variability  in
\sloang\  compared to \sloani\, or more  high-frequency variability in \sloani\
compared to \sloang. The latter would arise if there were more high frequency
variability in \sloani\ from the inner disk, such as from the  synchrotron
emission arising from the ADAF. 

The low/hard and quiescent states of the XRTs are often interpreted as  having
a truncated disc with an ADAF central region. An ADAF model has been
successfully applied to \target\ (\citealt{McClintock01b};  \citealt{Esin01};
\citealt{Chaty03}). With spectral coverage which includes X-rays, Extreme
Ultraviolet (EUV) and radio, model fits to the spectral energy distribution in
the low/hard state  suggests an ADAF inner radius of 350\,$\rm R_{\rm sch}$
\citep{Chaty03}.   Recently \citet{McClintock03a} have determined the 
multi-wavelength X-ray/UV/optical  quiescent spectrum  of \target\ and  
find that the spectrum  has two components  explained by an ADAF in the disc
interior  with a radius of   $\rm \sim 10^{4}\,R_{\rm sch}$, and an optical/UV
continuum that resembles a 13,000\,K disc  blackbody model spectrum  with a
radius of  1500\,$\rm R_{\rm sch}$ arising from the disc/stream impact region.

The exterior accretion disc truncated at a radius of   $\rm R_{\rm tr} \sim
10^{4}\,R_{\rm sch}$ is responsible for a substantial  fraction of the
optical/UV spectrum.   Although an ADAF of this large size  fits the X-ray data
it does not produce a significant  UV/optical  component \citep{Narayan97}. 
Furthermore, a significant ADAF contribution to the optical/UV emission is
ruled out  because of (1) the presence of broad emission lines; (2) the
Planckian shape of the   optical/UV continuum spectrum, and (3) the large
orbital  modulation of the NUV continuum which cannot be attributed to
synchrotron  emission from an ADAF. For the models outlined  in
sections\,\ref{BMODELS} and \ref{LFQPO} we also find a quiescent ADAF radius of
$\rm R_{\rm crit} \sim 10^{4}\,R_{\rm sch}$, comparable to the size of the
quiescent ADAF region in the black hole A0620--00 and V404\,Cyg, suggesting
that the  size of the ADAF region is determined by the same underlying
physical  mechanism, such as viscosity and inner disc temperature.

In section\,\ref{VAR} we determined the  blackbody colour and radius of the
short-term rapid flares. We find that  the flares have a blackbody temperature
of $\sim$3500\,K and a  radius of $\sim$0.10\,\Rsun (=3200\,$\rm R_{\rm sch}$),
which is larger and cooler  than the size and temperature of the gas
stream/disc impact region determined by \citet{McClintock03a}. Most probably
the short-term flares arise from {\it all} regions in  the outer parts of the
disc. However, it should be noted that \citet{Torres04} find no evidence for a
bright spot (disc/stream impact region) from their optical spectroscopy taken
in Jan 2003; a 13,000\,K spectrum \citep{McClintock03a} is not enough to
produce Balmer emission lines. It could be that the mass transfer from the
secondary star has reduced considerably and so the parameters for the
short-term rapid flares we derive could be consistent with an origin from the
bright spot. Only high-time resolution spectroscopic observations can resolve
this issue.

\subsection{Quiescent state models}

Although the quiescent thermal ADAF model, which predicts a curvature in the
X-ray spectrum  primarily because of the assumption of a thermal (power--law)
energy distribution for the  electrons (\citealt{Esin97}; \citealt{Esin98}), is
consistent with the observed X-ray data \citep{McClintock03a}, it should be
noted that given the quality of the data, a power--law model as predicted by an
ADAF with non-thermal electrons  cannot be ruled out. The quiescent  optical/UV
data suggests a 13,000\,K multi-colour blackbody spectrum, arising from the
accretion disc and/or where the gas stream impacts the disc.  In principle a
jet model, such as the one presented by \citet{Markoff01b} for \target\ in the
low/hard state, could be modified and applied to the  quiescent optical/X-ray
data shown in \citet{McClintock03a}. Since the jet model is a non-thermal
model, one would expect it to predict a power--law spectrum in the X-ray band,
which would be consistent with the data.  

In the context of a jet model,  it could be that the quiescent
infrared/optical  spectrum of \target\ is an extension of the  radio spectrum.
If the radio spectrum is due to self-absorbed synchrotron  emission from a
conical jet, then above some frequency (at which the whole jet is optically
thin) there should be a break to an  optically thin synchrotron spectrum.
Observations of the  low/hard state X-ray  source GX\,339--4 appears to have
identified such a break in the near-infrared $I$-band \citep{Corbel02}. For
GX\,339--4 the X-ray spectrum seems to lie on an extrapolation of the
non-thermal component. Looking at quiescent optical/X-ray SED of \target\
\citep{McClintock03a},  it seems difficult to describe the X-ray as a
continuation of the  optical data.  For \target\ a similar two-component
quiescent SED  overlapping in the optical/infrared,  with an additional
Planckian component due to an accretion disc may be expected, as is indeed
predicted by the jet models \citep{Markoff01b}, but it is clear that quiescent
data at radio and infrared wavelengths  are needed before any firm conclusions
can be drawn.

\section{Conclusion}

We present \uc\ observations of the  quiescent soft X-ray transient \target.
Superimposed on the \sloang\ and \sloani-band ellipsoidal lightcurves are rapid
flare events, the  PDS of which can be described by either a broken power--law
model with a break frequency at $\sim$2\,mHz or a   power--law model plus a
broad quasi-periodic oscillation (QPO) at $\sim$2\,mHz.  In either caase, the
size of the quiescent ADAF region  is esimated to be $\sim 10^{4}$
Schwarzschild radii, similar to that observed in other quiescent black hole
X-ray transients. The similarities between that the low/hard state  PDS
(80\,mHz QPO and a  $\sim$23\,mHz break-frequency) with the quiescent state
optical PDS  suggest a similar origin for the optical and X-ray variability,
most likely from regions at/near the ADAF region.

\section*{Acknowledgments}

TS and JC acknowledge support from the Spanish Ministry of Science  and
Technology  under the grant AYA\,2002\,03570 and the programme Ram\'{o}n y
Cajal. TRM acknowledges the support of a PPARC Senior Research Fellowship. 
ULTRACAM is supported by PPARC grant PPA/G/S/2002/00092.  Based on observations
made with  the William Herschel Telescope   operated on the island of La Palma
by the Issac Newton Group in the Spanish Observatorio del Roque de los
Muchachos of the Instituto de Astrof\'\i{}sica de Canarias.

\end{document}